\documentclass[aps,pre,twocolumn,groupaddress,showpacs,floatfix]{revtex4}

\usepackage{amsmath,amssymb,graphicx,color}

\begin{document}
\title{Work fluctuations for Bose particles in grand canonical initial states}

\author{Juyeon Yi}
\affiliation{Department of Physics, Pusan National University,
Busan 609-735, Korea}

\author{Yong Woon Kim}
\affiliation{Graduate School of Nanoscience and Technology, Korea Advanced Institute of Science and Technology, Daejeon 305-701, Korea}

\author{Peter Talkner}
\affiliation{Institute of Physics, University of Augsburg,
Universit\"{a}tsstrasse 1, D-86135, Augsburg, Germany}

\date{\today}

\begin{abstract}

We consider bosons in a harmonic trap and investigate the fluctuations of the work performed by an adiabatic change of the trap curvature. Depending on the reservoir conditions such as temperature and chemical potential that provide the initial equilibrium state, the exponentiated work average~(EWA) defined in the context of the Crooks relation and the Jarzynski equality may diverge if the trap becomes wider. We investigate how the probability distribution function~(PDF) of the work signals this divergence. It is shown that at low temperatures the PDF is highly asymmetric
with a steep fall off at one side and an exponential tail at the other side.
For high temperatures it is closer to a symmetric distribution approaching a Gaussian form.
These properties of the work PDF are discussed in relation to the convergence of the EWA and to the existence of the hypothetical equilibrium state
to which those thermodynamic potential changes refer that enter both the Crooks relation and the Jarzynski equality.
\end{abstract}


\maketitle
\section{Introduction}
In recent years fluctuation theorems that allow to infer equilibrium properties of a system from nonequilibrium processes
have attracted considerable attention. The Jarzynski equality~(JE)~\cite{Jarzynski} reading
\begin{equation}\label{JE}
\langle e^{-\beta w}\rangle =e^{-\beta \Delta F},
\end{equation}
provides a prominent example of these relations.
It relates the free energy change $\Delta F$ to the statistics of work $w$ that is performed by a time-dependent force acting on a system that initially is prepared in a canonical equilibrium state at the temperature $T=1/k_B \beta$. The weight with which the average of the exponentiated work is performed is determined by the probability density function (PDF) of work, $p(w)$. It represents the frequency of outcomes from independent runs of the same force protocol starting in equilibrium at the same temperature.
The free energy change $\Delta F$
gives the difference between free energies of the initial state and of a hypothetical thermal equilibrium state of the considered system at the  initial temperature $T$ with clamped forces at the values at the end of the force protocol.
First, the JE was found for classical systems and later on confirmed for quantum mechanical systems~~\cite{quantum1,quantum2,talkner1,talkner2,talkner3,EHM,CHT}.

A similar form of the JE holds for grand canonical initial states~(GCI) which allow both energy and particle number fluctuations. It then takes the form~\cite{seifert,qmnumber,YTC}:
\begin{equation}\label{gci}
\langle e^{-\beta w}e^{\beta \mu n}\rangle_{gc}=e^{-\beta \Delta \Phi},
\end{equation}
where $w$ and $\beta$ are defined as above. Further, $\mu$ denotes the chemical potential of the reservoir, and $n$ the difference of the particle numbers at the end and at the beginning of the force protocol. Similarly as the work $w$ also the particle number change $n$ is a random quantity. These random outcomes are described by a joint PDF $p(w,n)$.
Here $\Delta \Phi = \Phi_{f}-\Phi_{i}$ is given by the difference of the grand canonical potentials $\Phi_{i}$ and $\Phi_{f}$ which respectively correspond to the initial equilibrium state and to a hypothetical equilibrium state at the inverse temperature $\beta$ and chemical potential $\mu$ with clamped force values at the end of the force protocol.

While temperature is universally confined to positive values, the upper admissible bound of the chemical potential is system dependent. To illustrate this fact we consider a system of non-interacting identical particles for which the average particle number in a grand canonical potential at inverse temperature $\beta$ and chemical potential $\mu$ is determined by \cite{LL}
\begin{equation}
N_{av} = \sum_{\epsilon}\frac{1}{e^{\beta (\epsilon-\mu)}\mp 1},
\end{equation}
where the sum is performed over the single particle energy spectrum, and the $+$ and $-$ signs refer to
Fermi-Dirac and Bose-Einstein statistics, respectively. Considering the case of a single particle spectrum which is bounded from below by the ground state, 
for fermions, the $+$ sign guarantees the convergence of the sum for any real value of the chemical potential.
However, for a system made of bosons, this sum only converges if the chemical potential is smaller than the ground state energy.
With the divergence of the average particle number the grand canonical partition function diverges and accordingly 
the grand canonical potential becomes negatively divergent. In the context of the fluctuation theorem given by Eq.~(\ref{gci}), this
implies that the averaged exponentiated linear combination of work and number change, that is, the left hand side of Eq.~(\ref{gci}),
diverges for Bosonic systems under the action of protocols which lead to a lowering of the single particle ground state energy below the level of the chemical potential of the initial grand canonical equilibrium state.

The purpose of this work is to address this issue
of the fluctuation theorem for GCI, Eq.~(\ref{gci}).
For a concrete discussion, we consider non-interacting bosons residing in a three-dimensional harmonic potential.
Such systems have been treated as a theoretical models of trapped atomic gases undergoing Bose-Einstein condensation~\cite{becexp1,becexp2,becexp3}. Section II is devoted to a brief introduction to the considered model system and the required procedure for obtaining the work statistics.
In Sec. III, we present a symmetry relation of the Crooks-Tasaki type for GCI, and point out the condition 
of its existence. Section IV specifies the work protocol to be considered in this work. The analytic properties of the corresponding characteristic function are discussed in Sec. V. The PDFs for extreme temperature regimes are analytically obtained in Sec. VI. It is shown that the PDF at very low temperatures has a long tail being responsible for the divergence of the exponentiated work average~(EWA). In Sec. VII, the convergence of this average for general temperatures is examined by a numerical evaluation of the PDFs, and discussed in relation to the existence of the hypothetical equilibrium state.

\section{system}
We consider $N$ identical Bose particles of mass $m$ moving in a three-dimensional
symmetric harmonic trap with a curvature changing in time $t$. The governing Hamiltonian
at an instantaneous time $t$ reads
\begin{equation}\label{ham1}
{\cal H}(t)=\sum_{i=1}^{N}\left[
\frac{{\mathbf  p}_{i}^{2}}{2m}+\frac{1}{2}m\omega^{2}(t){\mathbf x}_{i}^{2}\right].
\end{equation}
This Hamiltonian is widely studied as a simple model of the Bose-Einstein condensation~\cite{shodos1,shodos2,becfnum}.
However, the feature of nonequilibrium work statistics to be drawn here is not specifically related to the condensation transition.
The single particle energy spectrum at time $t$ is
given by
\begin{equation}
\epsilon_{\ell}(t)=(\ell_{x}+\ell_{y}+\ell_{z}+3/2)\hbar\omega(t)
\label{eps}
\end{equation}
with a set of non-negative integers  $\ell\equiv (\ell_{x},\ell_{y},\ell_{z})$.
\subsection{Equilibrium properties}
This many particle system is supposed to initially stay in equilibrium with a reservoir having prescribed values of the chemical potential $\mu$ and the inverse temperature $\beta$.  For this initial equilibrium state the initial grand canonical partition function and the initial average number of particles are determined by
\begin{eqnarray}\label{avenum}
{\cal Q}_{i} &=&\prod_{\ell}[1-z e^{\beta\epsilon_{\ell}(0)}]^{-1} \\ \nonumber
N_{av} &=&
\sum_{\ell}\frac{1}{z^{-1}e^{\beta \epsilon_{\ell}(0)}-1}\:,
\end{eqnarray}
where $z=e^{\beta\mu}$ denotes the fugacity.
The condensation fraction is given by $N_{0}/N_{av}$ with $N_{0}$ denoting the
particle occupancy in the ground state,  $N_{0}=1/(z^{-1}e^{\beta \epsilon_{0}(0)}-1)$.
For sufficiently large $N_{av}$, the condensation curve falls
onto the critical line, $N_{0}/N_{av} =1-(T/T^{(0)}_{c})^{3}$ if $T \leq T_{c}^{(0)}$ and
$N_{0}/N_{av} =0$, otherwise, with the critical temperature $T_{c}^{(0)}=(N_{av}/\zeta(3))^{1/3}(\hbar\omega/k_{B})$, where $\zeta(3)\approx 1.202$.
When the number of particles is finite, the transition becomes smeared out and the critical temperature is modified as~\cite{becfnum}
\begin{equation}
T_{c}/T_{c}^{(0)}=(1-0.7275 N_{av}^{-1/3}) \:.
\end{equation}
In the sequel we shall use $T_c$ as temperature unit.

\begin{figure}[t]
\includegraphics[width=0.8\columnwidth]{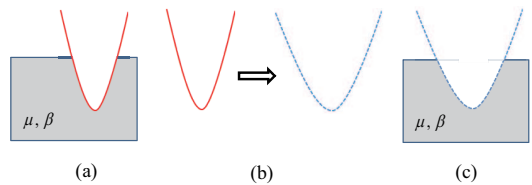}
 \caption{(a) Grand canonical initial states of  Bose particles in a harmonic potential that
  are in equilibrium with a reservoir of chemical potential $\mu$ and  inverse temperature $\beta$. (b) The system is decoupled from the reservoir and then work is performed by changing the curvature of the potential (expansion of the potential in the figure). (c) Hypothetical equilibrium at the same inverse temperature $\beta$ and chemical potential $\mu$
  with the harmonic potential curvature clamped at the end of the force protocol.}
\end{figure}

\subsection{Doing work}
We here sketch  a gedankenexperiment that elucidates the relevant steps implied by Eq.~(\ref{gci}). Fig.~1(a) depicts the initial
equilibrium state of the considered many particle system in weak 
contact with a reservoir that may exchange particles and energy with the system
controlled by the chemical potential $\mu$ and the inverse temperature $\beta$. The grand potential $\Phi_{i}=-k_{B}T\ln {\cal Q}_{i}$ for this initial state is determined by the reservoir parameters as well as by the microscopic details such as the initial curvature of the potential defining the oscillation frequency $\omega$. Once the system has approached
the grand canonical equilibrium state, it is decoupled from the reservoir, its energy and particle number are determined, and afterward the curvature of the trap is changed according to a designed protocol. Finally energy and particle number are again measured. The change of energy determines the work $w$ performed on the system in this particular realization. The work and the particle number change $n$ are finally registered. This procedure must be repeated many times, always starting from the same equilibrium state and following the same protocol such that the joint probability $p(w,n)$ can be estimated,
or, directly the exponential average $\langle e^{-\beta (w - \mu n)} \rangle$ can be estimated.
According to Eq.~(2) this average value 
coincides with the ratio of the two partition functions. The denominator is given by the partition function of the initial system and hence determined by the initial initial trap curvature as well as by $\beta$ and $\mu$. The numerator refers to the hypothetical equilibrium of system with the final trap curvature 
see Fig. 1~(c).


There is no restriction for the finally reached trap curvature. In particular, the trap may be widened to such an extend that the ground state energy falls below the chemical potential of the initial state. Then the hypothetical equilibrium state is not properly defined and formally leads to a divergent grand canonical partition function and a negative, divergent grand canonical potential. At the same time the exponential average of $-\beta (w- \mu n)$ also diverges.


\section{characteristic function}
Along with the JE, the Tasaki-Crooks relation ~\cite{quantum1, crooks} reading
\begin{equation}\label{CT}
e^{-\beta w} p(w)=e^{-\beta\Delta F}p_{b}(-w),
\end{equation}
provides a connection between the PDFs $p(w)$ of the original process and the PDF $p_b(w)$ of the backward process for systems initially prepared in a canonical equilibrium state.
Here the backward process starts at the hypothetical equilibrium state retracing the force protocol from its final to the initial value of the forward process.
From the correspondence between Eq.~(\ref{JE}) and Eq.~(\ref{gci}),
one may expect the existence of a Crooks-Tasaki relation for the GCI of the form
\begin{equation}\label{gtgci}
e^{-\beta(w-\mu n)}p(w,n)=e^{-\beta\Delta \Phi}p_{b}(-w,-n).
\end{equation}
Integrating over $w$ and summing over all possible values of $n$, we indeed obtain the generalized form of JE for the GCI, as given
in Eq.~(\ref{gci}).

The proof of this relation may be obtained in an analogous way as for the canonical case \cite{quantum1, talkner2,CHT} based on the characteristic function
$G(u,v)$. It is given as the Fourier transform of the joint PDF $p(w,n)$ with respect to both $w$ and $n$ and can be expressed as a two-time correlation function:
\begin{eqnarray}\label{invft}
G(u,v)&=&\sum_{n=-\infty}^{\infty}\int dw e^{iuw+ivn}p(w,n)  \\ \nonumber
&=&\langle e^{iu{\cal H}_{H}(\tau)+iv{\cal N}_{H}(\tau)}e^{-iu{\cal H}(0)-iv{\cal N}(0)}\rangle_{\rho_{i}},
\end{eqnarray}
where the average $\langle X\rangle_{\rho_{i}}=\mbox Tr X e^{-\beta {\cal H}(0)}e^{\beta \mu {\cal N}(0)}/{\cal Q}_{i}$ is performed over the initial grand canonical state
with ${\cal Q}_{i}$ being the grand canonical partition function of the initial state.
Here the index $H$ indicates operators in the Heisenberg picture given
by ${\cal O}_{H}(\tau)=U^{\dagger}(\tau,0){\cal O}(\tau)U(\tau,0)$.
Based on the micro-reversibility of the time evolution $U(\tau,t) = \Theta^\dagger U_b(t-\tau,0) \Theta $ relating the time evolution $U$ of the original process to the time evolution $U_b$ for the reversed protocol by means of the anti-unitary time-reversal operator $\Theta$\cite{CHT},  one obtains the following relation between characteristic functions of the forward and the backward process
\begin{equation}
{\cal Q}_{i}G(u,v)=G_{b}(-u+i\beta,-v-i\beta \mu){\cal Q}_{f}.
\end{equation}
Taking the inverse Fourier transform of this relation leads to Eq.~(\ref{gtgci}), where
$\beta\Delta\Phi = -(\ln {\cal Q}_{f}-\ln {\cal Q}_{i})$.

\section{protocol}
The protocol according to which the trap curvature is changed specifies the  time-dependent change of the frequency $\omega(t)$ within a time interval $[0,\tau]$. In the present investigation we assume that it consists in an adiabatically slow change connecting the boundary values
\begin{equation}
\omega(0)=\omega, \hspace{0.3cm} \omega(\tau)=(1+\gamma) \omega.
\end{equation}
A positive~(negative) value of $\gamma$ indicates that the system is compressed~(expanded) during the protocol. With the adiabatic variation of the frequency the occupation numbers $n_\ell$ of the $\ell$'s single particle eigenstates remain unchanged such that the time evolution operator takes the form
\begin{equation}\label{prot}
U(t,0)=\sum_{\{n_{\ell}\}}| \{n_{\ell}\},t\rangle \langle \{n_{\ell}\},0| \:,
\end{equation}
where $|\{n_{\ell}\},t\rangle = |n_{0},n_{1},\cdots; t\rangle$ with $\sum_\ell n_\ell = N$ is an eigenfunction of the $N$-particle Hamiltonian (\ref{ham1})
and hence a solution of
\begin{eqnarray}\label{spec}
{\cal H}(t)|\{n_{\ell}\},t\rangle &=& E(t)|\{n_{\ell}\},t\rangle \\ \nonumber
E(t)&=&\sum_{\ell}\epsilon_{\ell}(t)n_{\ell} \:.
\end{eqnarray}
The corresponding $N$-particle eigenvalue $E(t)$ is expressed in terms of the single-particle energy eigenvalues $\epsilon_\ell(t)$
given by Eq.~(\ref{eps}) and the occupation numbers $n_{\ell}$ of these states.
We consider this adiabatic protocol for the sake of simplicity. Although the shape of the work PDF will depend on the details of the specific protocol their relevant qualitative features leading to a diverging EWA are expected to be independent of those details.

\section{Analytic properties of the characteristic function}
Using Eqs.~(\ref{prot}) and (\ref{spec}), we obtain for the Hamiltonian and the number operator in the Heisenberg picture at the final time $\tau$ of the protocol
\begin{equation}
\begin{split}
{\cal H}_{H}(\tau)
&=\sum_{\{n_{\ell}\}}\!^{'}E(\tau)|\{n_{\ell}\},0\rangle \langle \{n_{\ell}\};0| \:, \\
{\cal N}_{H}(\tau)&= \sum_{\{n_{\ell}\}}\!^{'} n_{\ell}|\{n_{\ell}\},0\rangle \langle \{n_{\ell}\};0|={\cal N}(0),
\end{split}
\end{equation}
where $\sum_{\{n_{\ell}\}}^{'}$ denotes the summation under the constraint
$\sum_{\ell}n_{\ell}=N$. As a consequence of the number conservation for the considered protocol, the characteristic function becomes independent of the variable $v$ which is conjugate to the number change $n$. Hence, we get $G(u,v) =G(u)$, implying $p(w,n) = p(w) \delta_{n,0}$ for the joint probability.

Since all operators entering the characteristic function (\ref{invft}) under the trace are diagonal with respect to the eigenbasis of the initial Hamiltonian, all of them commute with each other and it therefore is 
straightforward to write
\begin{equation}\label{cha}
\begin{split}
G(u)&={\cal Q}_{i}^{-1}\sum_{N=0}^{\infty}\sum_{\{n_{\ell}\}}\!^{'}\prod_{\ell}
e^{iu\epsilon_{\ell}(\tau)n_{\ell}}
e^{-i(u- i \beta)\epsilon_{\ell}(0)n_{\ell}}z^{n_{\ell}}\\
&={\cal Q}_{i}^{-1}\prod_{l=0}^{\infty} \left [ 1- e^{(i u \gamma -  \beta)\hbar \omega (l+3/2)} \right ]^{-g(l)} \: .
\end{split}
\end{equation}
Here the product on the right hand side of the first line extends over the triple index $\ell =(\ell_x,\ell_y,\ell_z)$.
The restriction in the second sum on the right hand side of the first line is lifted by the first sum over $N$. Therefore all sums over the $n_\ell$s can be performed in closed form leading to the expression in the second line.
Because of the degeneracy of the single particle energies $\epsilon_\ell(t)$ having the same value for a given $l =\ell_x+\ell_y+\ell_z$, see  Eq.~(\ref{eps}), the product in the second line can be taken for $l=0,1,2,\ldots$. The degree of the degeneracy of the single particle energies (\ref{eps}) is given by $g(l) =(l+1)(l+2)/2$.

Since $u$ and $\gamma$ only enter in the combination $u\gamma$
the work PDF, which is given by the inverse Fourier transform of the characteristic
function through Eq.~(\ref{invft}), depends on $w$ and $\gamma$ in terms of the ratio $w/\gamma$:
\begin{equation}\label{uniPDF}
p(w)=\frac{1}{|\gamma|}f(w/\gamma),
\end{equation}
where $f(x)=(2\pi)^{-1}\int_{-\infty}^{\infty}d\xi e^{-ix\xi}G(\xi)$ with $\xi=u\gamma$. This leads to a
symmetry relation between the work PDFs for compression~($\gamma > 0$) and
expansion~($\gamma < 0$):
\begin{equation}\label{symPDF}
p(w)|_{\gamma > 0}=p(-w)|_{\gamma < 0} \:.
\end{equation}
In presenting numerical results of the PDFs, we only consider the expansion case for a specific value of $\gamma$. However, thanks to the relations, Eqs.~(\ref{uniPDF}) and (\ref{symPDF}), PDFs for other cases not shown here can be visualized.

It is worthwhile here to mention that the convergence of $\langle e^{-\beta w}\rangle_{gc}$ is determined by the structure of the singularities of the characteristic function. The poles of $G(u)$ are located along the imaginary axis in the complex plane of
$u=u'+iu''$, where $G(iu'')^{-1}=0$, yielding
\begin{equation}
u''_l=-\frac{\beta [\epsilon_l(0)-\mu]}{\epsilon_l(0) \gamma}, \quad l = 1,2,\ldots \:.
\end{equation}
If $\gamma > 0$, then all poles are located in the lower half-plane~($u''_l<0$) so that a divergence of $G(iu''_l)=\langle e^{-u''_l w}\rangle_{gc}$ can only occur 
in the unphysical regime of negative temperatures. On the other hand, for $\gamma <0$ all pole positions are at positive values, $u''_l > 0$.
If the inverse temperature lies below the smallest pole position $u''_0= \beta  (\epsilon_l(0)-\mu)/(\epsilon_l(0) |\gamma|)$ the EWA $\langle e^{-\beta w}\rangle_{gc}$ is finite.
The opposite case leads to a divergent EWA.
Hence the condition for a finite EWA becomes
\begin{equation}\label{instability}
(1+\gamma) \epsilon_{0}(0) > \mu \:.
\end{equation}
Since $(1+\gamma) \epsilon_0(0) = \epsilon_0(\tau)$ is the single particle ground state in the trap at the end of the protocol this condition is identical with the condition of the existence of the hypothetical equilibrium state as explained at the end of Sec. II.

So far we have considered the chemical potential and the temperature as independent thermodynamic variables characterizing the initial state. In many practical applications it is more convenient to consider the average particle number as prescribed instead of the chemical potential. As a consequence the chemical potential then becomes a function of the average particle number and temperature and also the existence of a finite EWA then depends on temperature.

\section{asymptotic results}
In the extreme temperature limits, analytic forms of the work PDF can be obtained.
\subsection{High temperatures}
At high temperatures, the ground state energy is much smaller than the thermal energy, $\epsilon_{o} \ll k_{B}T$ and hence $\beta \epsilon_0 $ serves as expansion parameter.
The mean number of the particles and the initial grand partition function then are approximately given by
\begin{eqnarray}\label{hightempeq}
N_{av}
&\approx & z\sum_{l=0}^\infty g(l)e^{-\beta \epsilon_l(0)} \approx \frac{e^{\beta \mu}}{(\beta \epsilon_0)^3} \;, \\ \nonumber
\ln {\cal Q}_{i}&=&-\sum_{l=1}^\infty g(l)\ln[1-ze^{-\beta \epsilon_l(0)}]\approx N_{av}.
\end{eqnarray}
The second line gives the equation of states of an ideal gas. In the high temperature regime, positively or negatively large values of the work have highest probability for widening or narrowing, respectively, the trap. It is therefore sufficient to consider the contributions of the small values of $u$ to the characteristic function  yielding:
\begin{equation}
\ln G_\text{ht}(u) =  \frac{e^{\beta \mu}}{(\beta \hbar \omega)^3}\left [-1+ \frac{1}{(1- i \gamma u/ \beta)^3} \right ] \; .
\label{Ggcf}
\end{equation}
This asymptotic high temperature result agrees with the characteristic function of work for a classical system of non-interacting particles in a harmonic trap that initially stays in equilibrium with a reservoir and then experiences an adiabatic change of the trap curvature, see the appendix.
The PDF corresponding to the high temperature limit characteristic function is not known analytically. However, in the case of small curvature changes, i.e. $|\gamma| \ll 1$ one can perform an expansion in powers of $\gamma$, leading to:
\begin{equation}\label{chahightemp}
\ln G_\text{ht}(u)\approx \frac{e^{\beta \mu}}{(\beta \hbar \omega)^3}(3i\gamma u/\beta-6\gamma^{2}u^{2}/\beta^{2}).
\end{equation}
Within this approximation the work average and its standard deviation become:
\begin{eqnarray}
\langle w\rangle =  3N_{av}\gamma/\beta, \\ \nonumber
\sigma_{w}^{2}=4\gamma\langle w\rangle/\beta \:.
\end{eqnarray}
The corresponding Gaussian distribution function of the work is then given by
\begin{equation}\label{hightempPDF}
p(w)=\frac{1}{\sqrt{2\pi\sigma_{w}^{2}}}
\exp\left[-\frac{(w-\langle w\rangle)^{2}}{2\sigma_{w}^{2}}\right].
\end{equation}
As shown in Fig.~2(a), this PDF obtained for high temperatures and small curvature deformations
is in good agreement with the numerical evaluation of the PDF to be detailed in the next section.
In the high temperature approximation, (\ref{hightempPDF}) the EWA becomes
\begin{equation}
\ln\langle e^{-\beta w}\rangle_{gc} \approx -3N_{av}\gamma(1-2\gamma)\:.
\end{equation}
On the other hand, the grand canonical partition function of the hypothetical equilibrium state reads up to the second
order in $\gamma$
\[
\begin{split}
\ln{\cal Q}_{f}&= -\sum_{l}g(l)\ln[1-ze^{-\beta (1+\gamma) (l+\epsilon_{0})}] \\
& \approx N_{av}(1-3\gamma+6\gamma^{2}).
\end{split}
\]
This together with ${\cal Q}_{i}$ in Eq.~(\ref{hightempeq}) leads to
 $\ln[{\cal Q}_{f}/{\cal Q}_{i}]=\langle e^{-\beta w}\rangle_{gc}$, validating the Jarzynski equality, Eq.~(\ref{gci}),
 within the Gaussian approximation. In passing we note that this need not be expected since the Gaussian approximation often fails to describe the wings of the work distribution with sufficient accuracy to conform with the Jarzynski equality \cite{YT}.

\begin{figure}[t]
\resizebox{9cm}{!}{\includegraphics{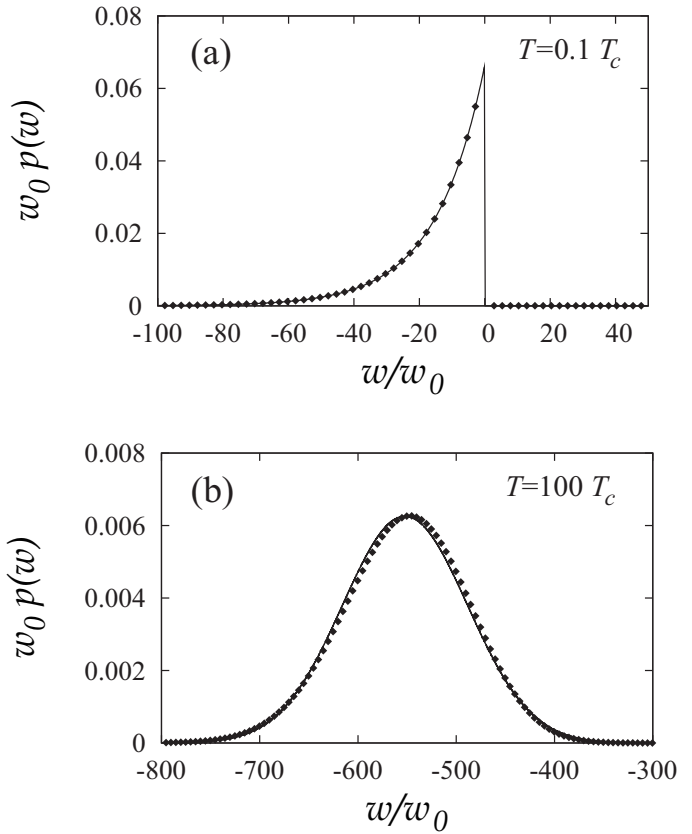}}
 \caption{Probability density functions of the work (in units of $w_{0}\equiv \hbar \omega$) for $\gamma =-0.1$ and $N_{av} =100$:
 (a) High temperature behavior at $T = 100T_{c}$, where the line is given by the analytic form, Eq.~(\ref{hightempPDF}). The panel (b) displays the low temperature behavior at $T=0.1 T_{c}$. The 
analytic result 
of Eq.~(\ref{lowtempPDF}) is depicted by the solid line. For comparison, we present the numerical results~(points) obtained by using Eqs.~(\ref{invft}) and (\ref{cha}), which are well consistent with the analytic forms
in the respective 
temperature limits.}
\end{figure}
\subsection{Low temperatures}
At sufficiently low temperatures the behavior of the system is determined by the ground state energy level~($l=0$).
The characteristic function, Eq.~(\ref{cha}) can be written as
 \begin{equation}
 G(u)\approx \frac{1-ze^{-\beta\epsilon_{0}(0)}}{1-ze^{-\beta\epsilon_{0}(0)}e^{iu\gamma\epsilon_{0}(0)}}\:,
 \end{equation}
which yields a distribution of work values $w_n = \gamma \epsilon_0(0) n$ with probabilities $p_n$ given by
\begin{equation}
p_n = (1- z e^{-\beta \epsilon_0(0)})  z^n e^{-\beta \epsilon_0(0) n} \:.
\label{lowtempPDFdis}
\end{equation}

For the EWA one obtains from this work distribution
\begin{equation}
\langle e^{-\beta w} \rangle = (1-e^{-\beta(\epsilon_0(0) - \mu)})/(1-e^{-\beta(\epsilon_0(\tau) - \mu)})\:.
\label{ttja}
\end{equation}
The right hand side of this equation coincides with the ratio of the grand canonical partition functions (\ref{avenum}) at the end and the beginning of the protocol, each of which being evaluated at sufficiently low temperatures such that other than the ground state contributions can be neglected.
Hence, the Jarzynski equality also holds for the approximate low temperature work PDF (\ref{lowtempPDFdis}).
Note that for protocols leading to a ground state $\epsilon_0(\tau)$ less than the chemical potential  $\mu$ formally leads to the nonsensical result of a negative EWA, indicating the actual divergence of the sum representing this average.
Finally we note that for small values of the deformation parameter $\gamma$ the spacing between the allowed values of the work becomes smaller suggesting to approximate the discrete work distribution by a continuous PDF which can be written as a generalized exponential PDF
\begin{equation}\label{lowtempPDF}
p(w)=\frac{1}{|\langle w\rangle |}e^{-w/\langle w\rangle}\Theta(\gamma w).
\end{equation}
where $\Theta(x)$ denotes the Heaviside step function. The average work $\langle w \rangle$ follows from Eq. (\ref{lowtempPDFdis}) as
\begin{equation}
\langle w \rangle = \gamma \epsilon_0(0) \frac{z e^{-\beta \epsilon_0(0)}}{1-z e^{-\beta \epsilon_0(0)}}\:.
\label{ltaw}
\end{equation}
The sign of the average work is determined by that of $\gamma$: Compressing  the trap leads to positive, widening to negative work.
In accordance with Eq.~(\ref{instability}) there is no restriction for the existence of the EWA in the compression case. For widening though the characteristic scale on which the exponential distribution decays must be small enough that the increase of the exponentiated work $e^{\beta w}$ for negative $w$ is overcompensated  and a finite EWA exists. The quantitative condition $\beta |\langle w \rangle| <1$ following from (\ref{ltaw}) is identical with the condition implied by the existence of the hypothetical equilibrium state, $\epsilon_0(0) > \mu$.

\begin{figure}[t]
\resizebox{9cm}{!}{\includegraphics{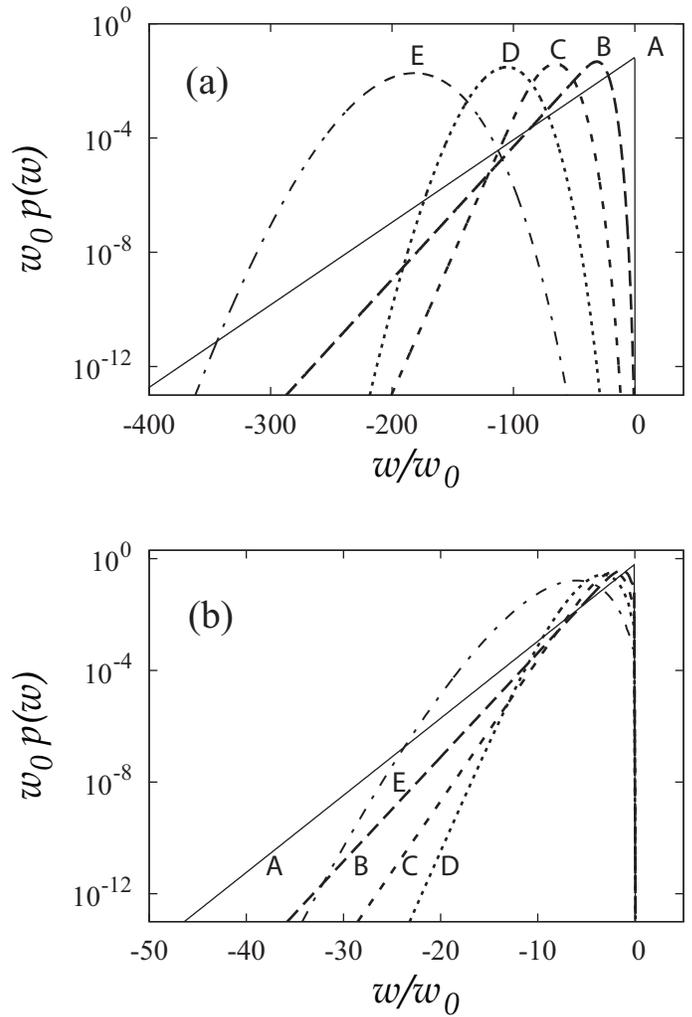}}
 \caption{
 Probability distribution functions of the work (in units of $w_{0}\equiv \hbar \omega$) for the potential expansion $\gamma =-0.1$: The panel (a) and (b) are for $N_{av}=100$ and
 $N_{av}=10$, respectively. The curves in each panels are obtained
 for various temperatures $T/T_c =0.1$ (A), 0.7 (B), 0.9 (C), 1.1 (D), and 1.7 (E), which correspond to the crosses in Fig.~4 (a) below. At the relatively high
temperature, the PDF for $N_{av}=100$ is approximately Gaussian.
With decreasing temperature, the asymmetry of the PDF becomes more pronounced
in the form of a heavier
tail in the 
negative work region, and finally
converges into an exponential PDF~(see the curve A). For a system
of smaller number of particles, the PDF 
is more asymmetric already at higher temperatures and becomes even more skewed at low temperatures.}
\end{figure}

\section{Numerical results}
In order to investigate the behavior of the work PDFs in the intermediate temperature regime,
we numerically evaluated the characteristic function in Eq.~(\ref{cha}) and obtained the work PDF
by means of an inverse transformation of Eq.~(\ref{invft}).
For that purpose, we used the fast Fourier transform algorithm proposed
by Danielson and Lanczos~\cite{nrc}.
Figure 3(a) displays the PDFs for $\gamma =0.9$ and $N_{av} =100$ at  various temperatures.
At $T=1.7T_{c}$~(see the curve labeled by E), the work PDF exhibits a 
decay that, on the logarithm scale, is faster than linear.
At lower temperatures, the PDF becomes negatively skewed developing a more pronounced tail
in the region of large negative work. At the extremely low temperature,
$T=0.1T_{c}$~(
labeled by A in Fig. 3), the PDF approaches the generalized exponential distribution, Eq.~(\ref{lowtempPDF}). The overall
feature of this temperature dependence is confirmed also for smaller average
particle numbers  $N_{av}=10$, the PDFs of which are shown in the panel (b).

\begin{figure}[b]
\resizebox{9cm}{!}{\includegraphics{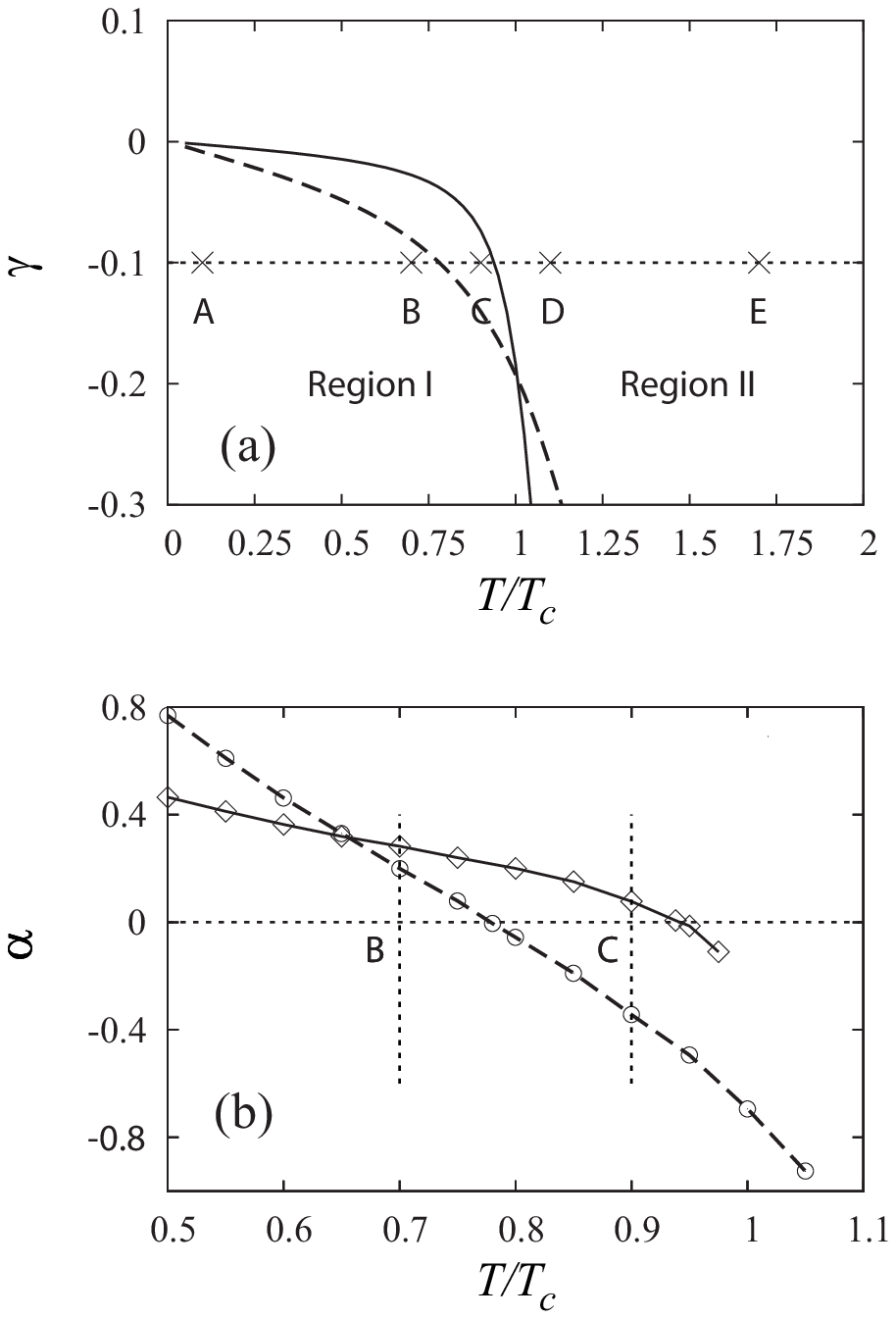}}
\caption{(a) Phase diagram depicting the region of existence of the hypothetical grand canonical state
in the $\gamma - T/T_c$ plane. The borders separating the two regions of existence (II) and nonexistence (I) are depicted by a solid line for $N_{av}=100$ and by a dotted line for $N_{av}=10$. Compression ($\gamma > 0$) belongs to
the region II irrespective of the temperature.
If the potential is expanded the hypothetical equilibrium ceases to exist and region I is entered provided the temperature is low enough. At the expansion factor $\gamma = -0.1$~(dotted horizontal line), the region I is entered
at $T \approx 0.95 T_c$ for $N_{av}=100$ and at $T\approx  0.77 T_c$ for $N_{av}=10$.
The points A, B, C, D and E (crosses) correspond to the accordingly marked work PDFs that are displayed in Fig.~3. The convergence measure $\alpha$ introduced in Eq. (\ref{alpha}) is displayed in panel (b) as a function of $T/T_c$ for $\gamma =-0.1$ and for $N_{av}=100$ ($\diamond$) and $N_{av}=10$~($\circ$).
The temperature values at which $\alpha$ changes the sign indicate the transition between finite and divergent EWA in agreement with the corresponding temperature values read off from panel (a). 
The vertical lines refer 
to the temperatures of 
the points B and C. For $N_{av}=10$, 
the crossing points with the lines corresponding to $B$ and $C$ give positive and negative $\alpha$ values, respectively, in accordance with panel (a).
For $N_{av}=100$ 
both points yield positive 
$\alpha$ values in agreement with panel (a).}
\end{figure}

As mentioned, the divergence of the EWA sets in when the ground state energy at the end of the work protocol is identical
to the chemical potential of the reservoir. Figure 4 (a) displays the critical line determined by $\gamma_{c}\epsilon_{0}=\mu-\epsilon_{0}$ for a given average number of particles.
The hypothetical equilibrium state exists only in the region II ($\gamma > \gamma_{c}$) of Fig.~4 (a).
This is the case when the work is done by compressing the potential~($\gamma >0$) but also in the limit of high temperatures.
On the other hand, upon expanding the potential 
the EWA diverges at low temperatures.
The key signature of this divergence
is reflected in the tail of the PDF at negative work values. This property of the PDF can be conveniently quantified by the parameter
\begin{equation}
\alpha = -\beta + \left(\frac{\partial \ln p(w)}{\partial w}\right)_{w=w_{c}},
\label{alpha}
\end{equation}
which determines the convergence rate of the integral $\int_{w_{c}}^{\infty}dw e^{-\beta w}p(w)$.
In our numerical investigation we
chose $w_{c}$ as the negative work for which the probability reaches the smallest possible value $p(w_{c})=10^{-13}$ 
within the numerical precision of our calculations.
Fig.~4(b) displays the temperature dependence of the convergence factor $\alpha$. In presenting the results, we show only the $\alpha$ values for PDFs whose negative work tails  approach
an exponential behavior such that a reliable value of $\alpha$ results. In the high temperature region the $\alpha$ values are negative.
Positive $\alpha$ values which occur in the low temperature regime
indicate the divergence of $\langle e^{-\beta w}\rangle_{gc}$. The regions of positive $\alpha$ indeed coincide with the instability
regions given in Fig.~4(a).

\section{summary}
We 
studied a subtlety of fluctuation theorems 
specific for Bose particles which initially are prepared in a grand canonical equilibrium state. When the considered protocol finally leads to
a Hamiltonian whose ground state energy per particle is less than the chemical potential of the initial state then
the hypothetical equilibrium state is ill defined and its corresponding grand canonical partition function and grand potential do not exist.
For the statistics of work fluctuations this means that the exponential work average diverges in spite of the fact that the moments of the work of all orders are finite. In this situation the Jarzynski equality looses its meaning
and also the Tasaki Crooks relation becomes pointless because the initial equilibrium state for the backward process does not exist and therefore cannot be prepared.

In order to illustrate this issue, we considered Bose particles in a three dimensional harmonic trap and investigated the statistics of the work done by changing the trap curvature adiabatically.
The probability distribution of the work at low temperatures follows a generalized exponential distribution which
 has a more pronounced tail than the Gaussian work distribution which is valid at high temperatures.
We presented analytic forms of the PDFs in the extreme temperature regimes, which are in good agreement with numerical results. In the intermediate regime the numerical results illustrate the transition between the extreme temperature cases.
As a quantitative measure for the decay of the PDFs
we examined the decay rate $\alpha$ the sign of which governs the convergence of $\langle e^{-\beta w}\rangle$. When the work is done by expanding the trap, $\alpha$ is always negative to guarantee the convergence. On the other hand, when the trap is compressed, $\alpha$ undergoes a sign change upon varying the temperature at constant average particle numbers. 
Hence, in the  low temperature regime  the average $\langle e^{-\beta w}\rangle_{gc}$ diverges.

Y.W.K. acknowledges support from Basic Science Research Program through the National Research Foundation of Korea (NRF) funded by the Ministry
of Education, Science and Technology, Korea (Grant No. 2010-0025196).

\appendix
\section{Classical Limit}
If the Hamiltonian ${\cal H}(t)$ commutes with $\cal{N}$ for all times during the protocol the particle number is conserved, and the characteristic function, Eq.~(\ref{invft}), can be decomposed into $N$-particle components:
\begin{equation}\label{GgcGc}
\begin{split}
G(u) &= {\cal Q}_{i}^{-1}\sum_N e^{\beta \mu N} \text{Tr}_N e^{i u {\cal H}_H(\tau)} e^{-i u {\cal H}} e^{- \beta H}\\
&= {\cal Q}_{i}^{-1}\sum_N e^{\beta \mu N} Z_N g_N(u) \:,
\end{split}
\end{equation}
where $Z_N$ is the canonical partition function of the $N$-particle system, and $g_{N}(u)$ is the characteristic function of work for the $N$-particle system with initial
canonical equilibrium state. In the particular case of non-interacting Boltzmann particles, the canonical $N$ particle partition function can be expressed by the single-particle partition function $Z_s$ as $Z_{N}=(Z_{s})^{N}/N!$ similarly the canonical N-particle generating function in terms of the one-particle generating function $g_s(u)$ as $g_{N}(u)=[g_{s}(u)]^{N}$. 
Using ${\cal Q}_{i}=\sum_{N}e^{\beta\mu N}Z_{N} = \exp[e^{\beta \mu}Z_{s}]$
and summing up the series, one obtains the characteristic function for the classical particles,
\begin{equation}\label{Gc1}
G_{c}(u) = \exp \left [ e^{\beta \mu} Z_s (g_s(u)-1) \right ]\:.
\end{equation}
For the example of particles subject to a three-dimensional isotropic harmonic potential undergoing a change of its curvature, 
the single-particle characteristic
function becomes
\[
g_s(u) = \left\{\int \frac{dp dq}{Z_{s}h} e^{iu [H(p(\tau),q(\tau),\tau) - H(p,q,0)]} e^{-\beta H(p,q,0)}\right\}^{3}\:,
\]
where $h$ is Planck's constant. In the particular case of an adiabatically slow change of the potential curvature 
the time dependent Hamiltonian  can be expressed in terms of the action $I$ 
to yield
\begin{equation}\label{H}
H(p(t),q(t),t) = \frac{1}{2m} p(t)^2 + \frac{m}{2} \omega^2(t) q(t)^2 = \omega(t) I,
\end{equation}
With the invariance of the action under adiabatic changes we get
\[
H(p(\tau),q(\tau),\tau) - H(p,q,0) = (\omega(\tau) - \omega(0)) I = \gamma \omega I .
\]
Combined with $\int dp dq = 2\pi \int_0^\infty dI$, this gives
\begin{equation}\label{ZG}
\begin{split}
Z_s g_s(u) &= \hbar^{-3} \left[\int_0^\infty d I e^{-(\beta - i \gamma u)\omega I}\right]^{3}\\
&=\frac{1}{(\beta -i \gamma u)^{3}(\hbar \omega)^{3}} \:,
\end{split}
\end{equation}
and $Z_{s}=1/(\beta\hbar \omega)^{3}$.
Hence we find for the grand canonical characteristic function
\begin{equation}
G_{c}(u) =\exp \left \{\frac{e^{\beta \mu}}{(\beta \hbar \omega)^{3}} \left[\frac{1}{(1- i \gamma u/ \beta)^{3}}-1 \right] \right \},
\end{equation}
which coincides with the quantum expression of characteristic function at high temperatures, Eq.~(\ref{Ggcf}).

\end{document}